# Testing CCC+TL Cosmology with Galaxy Rotation Curves

Rajendra P. Gupta

Department of Physics, University of Ottawa, Ottawa, ON K1N 6N5, Canada; rgupta4@uottawa.ca

**Abstract**

This paper aims to explore whether *astrophysical* observations, primarily galaxy rotation curves, result from covarying coupling constants (CCC) rather than from dark matter. We have shown in earlier papers that *cosmological* observations, such as supernovae type 1a (Pantheon+), the small size of galaxies at cosmic dawn, baryon acoustic oscillations (BAO), the sound horizon in the cosmic microwave background (CMB), and time dilation effect, can be easily accounted for without requiring dark energy and dark matter when coupling constants are permitted to evolve in an expanding Universe, as predicted by Dirac, and the redshift is considered jointly due to the Universe's expansion and Zwicky's tired light (TL) effect. Here, we show that the CCC parameter $\alpha$ is responsible for generating the illusion of dark matter and dark energy, which we call $\alpha$-matter and $\alpha$-energy, and is influenced by the baryonic matter density distribution. While cosmologically $\alpha$ is a constant determined for the homogenous and *isotropic* Universe, e.g., by fitting Pantheon+ data, it can vary locally due to the extreme *anisotropy* of the matter distribution. Thus, in high baryonic density regions, one expects $\alpha$-matter and $\alpha$-energy densities to be relatively low and vice versa. We present its application to a few galaxy rotation curves from the SPARC database and find the results promising.

## 1. Introduction

Galaxy rotation curves present one of the most substantial pieces of *astrophysical* evidence for the presence of dark matter in the Universe. It is different from the *cosmological* evidence based on the Friedmann equations fitting the observations, such as supernovae type 1a standard candle data and the CMB anisotropy power spectrum. Here, cosmological means studying the Universe parametrized, assuming it is isotropic and homogenous, whereas astrophysical means gravitationally bound virialized systems isolated from the Hubble flow. Several alternatives have been presented for cosmological or astrophysical dark matter, but none that are consistent with both, especially since all attempts to date have not been successful in finding dark matter at the elementary or macroscopic particle level [e.g., 1-10]. Macroscopic particles include massive compact halo objects (MACHOs), e.g., neutron stars, white dwarfs, brown dwarfs, and planets [11]. Even if a dark matter particle is found, such particle matter must be six times the baryon mass to be cosmologically relevant.

While the history of rotation curves can be traced back to Slipher (1914) [12], Page (1952) [13] and Burbidge and Burbidge (1960) [14] launched the modern approach to optical observations of rotation velocities in spiral galaxies. Excellent literature exists delineating the evolution of measurement techniques over decades of improved instrumentation that is explained by assuming the existence of dark matter due to the observed flatness of the galaxy's rotation curves to large radii [e.g., 15-23]. These and other studies finally resulted in a robust database, SPARC (*Spitzer* Photometry and Accurate Rotation Curves), for the rotation curves of 175 nearly disk galaxies with new surface photometry at 3.6 μm and high-quality rotation curves from previous H I/Hα studies. The database covers S0 to Irr morphologies, a ~5 dex luminosity range, and ~4 dex surface brightness range [23]. We have used a sample of galaxies from SPARC in our current work.

These studies are based on galaxies at low redshifts below about 0.1. Recent observations on spiral galaxies at redshifts ranging from 0.6 to 2.6 for a stack of ~100 galaxies show that the rotation curves fall off with the radius unlike the galaxies at low redshifts [2,3,5], although such a conclusion is debated by Tiley et al. [24] and Sharma et al.[25], based on the review of the method used by other authors and additional observations. Teklu et al. [26] showed that declining rotation curves and low dark matter in rotation-dominated galaxies at $z = 2$ appear naturally within the ΛCDM cosmology. Using gas kinematics, Fei et al.[27] assessed the dark matter content of two galaxies at $z = 6$ that are hosting quasars and estimated relatively large dark matter fractions within their effective radii, larger than those extrapolated from low redshift studies. One may, therefore, conclude that the flatness of galaxy rotation curves is not universal. However, the rotation curves at large distances suffer from limitations on telescope resolution. The further a galaxy is, the less of its dimmer outsides can be detected. Also, only the biggest and brightest galaxies can be seen at a great distance; see figures in Hofmeister et al.[28]. So, there is a systematic bias that needs to be addressed when these distant galaxies are analyzed [e.g., 29] and classified based on their kinematic properties [e.g., 30].

MOND (Modified Newtonian Dynamics) theory is the best-known alternative to dark matter. MOND, proposed by Milgrom in 1983 [31], acts in very low acceleration domains to account for the flat rotation curves observable in galaxies [see also 32-35]. He has interpreted it recently as a theory of modified inertia [36]. MOND belongs to the category of modified gravitational theories. Other such theories replacing dark matter and dark energy include the relativistic generalization of MOND in the form of tensor-vector-scalar gravity (TeVeS) [37], f(R) gravity [38], negative mass, dark fluid [39], entropic gravity [40], dark dimension [e.g., 41], retarded gravity [e.g., 42-44], and others [e.g., 45, 46]. The criticism that such theories are not compliant with the CMB anisotropy and matter power spectra observations has been refuted by Skordis and Złośnik [47]. Roberts et al. [48] used the gravothermal collapse of self-interacting dark matter to explain several galaxy rotation curves considered outliers in earlier studies. Mistele et al. [49] analyzed the gravitational potential derived from the gravitational lensing of isolated galaxies that imply rotation curves remain flat to a few hundred kpc [50] and showed that such curves may extend to the Mpc scale. These findings may imply that the galaxy halo is in thermal equilibrium even at larger radii, where particles do not have time to relax [51]. The universality of dark matter has been challenged by a new class of dark matter-free dwarf galaxies, raising concerns about galaxy formation models within the ΛCDM paradigm [52]. Here, we do not wish to provide a comprehensive literature review; it has been performed in several papers cited above.

It is worth mentioning the Special Issue of *Galaxies: Debate on the Physics of Galactic Rotation and the Existence of Dark Matter*, edited by Hofmeister and Criss in 2020 [10]. It covers the diversity of approaches to study Galactic rotations. Three papers discuss how the geometrical aspects of galaxies can reproduce galactic rotations without modifying gravitational dynamics or requiring dark matter. The geometries considered are the equatorial plane [53]; finite thickness disk [54, 55]; and the oblate [9, 10]. Sofue in 2020 reviewed the current status of the study on the rotation curve of the Milky Way and presented a unified rotation curve from the Galactic center to the galacto-centric distance of about 100 kpc. McGaugh [56] provided a review of a priori predictions made for the dynamics of rotating galaxies focusing on MOND. Marr [57] studied entropy and mass distribution in disk galaxies to understand the galaxy rotations.

Our attempt has been to explore if the covarying coupling constant (CCC) model of cosmology [58] is able to eliminate the need for dark matter to account for *astrophysical* observations as well as *cosmological* observations [59, 60]. We have already shown the hybrid model comprising the CCC model [61, 62] and the tired light (TL) phenomenon, the CCC+TL model, does not require dark matter to fit the Pantheon+ supernovae type 1a data and to resolve the so-called 'impossible early galaxy problem' [58]. Additionally, it is consistent with the baryon acoustic oscillation data, the CMB sound horizon angular size [59], time dilation observations, galaxy formation time scales, etc. [60].

Our focus here is to study the galaxy rotation curves within the CCC paradigm while leaving its application to galaxy cluster dynamics and gravitational lensing for a future paper. This paper is organized to present theoretical background in Section 2, galaxy rotation curves in Section 3, discussion in Section 4, and conclusions in Section 5.

## 2. Theoretical Background

In cosmology, the scale factor $a(t)$ accounts for the expansion of the Universe. We introduced another scale factor, $f(t)$, to consider the possibility of the length dimension of constants, i.e., the length unit itself evolving with the expansion. Then the speed of light evolves as $c(t) \sim f(t)$, gravitational constant as $G(t) \sim f(t)^3$, Planck constant as $h(t) \sim f(t)^2$, and Boltzmann constant as $k(t) \sim f(t)^2$, i.e., $G \sim c^3 \sim h^{3/2} \sim k^{3/2}$. This relationship can also be shown by local energy conservation law [61]. It modifies the FLRW metric and thus the Einstein equations, resulting in Friedmann equations slightly different from their familiar form. The modified FLRW metric, incorporating the covarying coupling constant (CCC) concept, is [58]

$$ds^2 = c_0^2 dt^2 f(t)^2 - a(t)^2 f(t)^2 \left( \frac{dr^2}{1-\kappa r^2} + r^2(d\theta^2 + \sin^2\theta \, d\phi^2) \right), \tag{1}$$

the modified Friedmann equations are

$$\left( \frac{\dot{a}}{a} + \alpha \right)^2 = \frac{8\pi G_0}{3c_0^2} \varepsilon - \frac{\kappa c_0^2}{a^2}, \text{ and} \tag{2}$$

$$\frac{\ddot{a}}{a} = -\frac{4\pi G_0}{3c_0^2}(\varepsilon + 3p) - \alpha \left( \frac{\dot{a}}{a} \right), \tag{3}$$

and the modified continuity equation is

$$\dot{\varepsilon} + 3\frac{\dot{a}}{a}(\varepsilon + p) = -\alpha(\varepsilon + 3p). \tag{4}$$

Here, $a$ is the scale factor (related to the redshift $z$ through $a = (1+z)^{-1}$), $G_0$ is the current value of the gravitational constant, $c_0$ is the current value of the speed of light, $\kappa$ is the curvature constant, $\alpha$ is a constant defining the variation of the constants through a function $f(t) = \exp(\alpha(t - t_0))$, with the cosmic time $t$ measured from the beginning of the Universe and $t_0$ being the current time, $\varepsilon$ is the energy density of all the components, and $p$ is their pressure[1]. Using the function $f(t)$, $c(t) = c_0 f(t)$, and $G = G_0 f(t)^3$ in the CCC model, the solution for Equation (4) for matter ($p = 0$) and radiation ($p = \varepsilon/3$) are, respectively,

$$\varepsilon_m = \varepsilon_{m,0} a^{-3} f^{-1}, \text{ and } \varepsilon_r = \varepsilon_{r,0} a^{-4} f^{-2}. \tag{5}$$

---

[1] It should be mentioned that the metric, Equation (1), can be transformed into the standard FLRW metric by redefining $t \to t' = (1/\alpha) \exp[\alpha(t - t_0)]$ and a new scale factor $a'(t) = a(t)f(t)$, thus yielding the standard Einstein tensor. However, when we write the complete Einstein equations [58], we find that we cannot eliminate $f(t)$ from the right-hand side, i.e., the right-hand side cannot be transformed into the standard form.

Defining the Hubble expansion parameter as $H \equiv \dot{a}/a$, we may write Equation (2) for a flat Universe ($\kappa = 0$) as

$$(H + \alpha)^2 = \frac{8\pi G_0}{3c_0^2} \varepsilon \Rightarrow \varepsilon_{c,0}^C \equiv \frac{3c_0^2 (H_0 + \alpha)^2}{8\pi G_0}. \tag{6}$$

This equation defines the current critical density $\varepsilon_{c,0}^C$ of the Universe in the CCC model that depends not only on the Hubble constant but also on the constant $\alpha$. Using Equations (5) and (6), we may write

$$(H + \alpha)^2 = (H_0 + \alpha)^2 \left( \Omega_{m,0} a^{-3} f^{-1} + \Omega_{r,0} a^{-4} f^{-2} \right). \tag{7}$$

In this equation, relative matter density is $\Omega_{m,0} \equiv \varepsilon_{m,0}/\varepsilon_{c,0}^C$ and relative radiation density is $\Omega_{r,0} \equiv \varepsilon_{r,0}/\varepsilon_{c,0}^C$. Since $\Omega_{r,0} \ll \Omega_{m,0}$, and since we do not have to worry about the dark energy density in the CCC model, Equation (7) simplifies to [60]

$$(H + \alpha)^2 = (H_0 + \alpha)^2 \left( a^{-3} f^{-1} + \Omega_{r,0} a^{-4} f^{-2} \right). \tag{8}$$

However, we may expand and rewrite Equation (2) as

$$\begin{aligned} H^2 &= \frac{8\pi G_0}{3c_0^2} \varepsilon - \alpha^2 - 2\alpha H \\ &= \frac{8\pi G_0}{3c_0^2} \varepsilon - \alpha^2 - 2\alpha \left( \left( \frac{8\pi G_0}{3c_0^2} \varepsilon \right)^{1/2} - \alpha \right) \\ &= \frac{8\pi G_0}{3c_0^2} \varepsilon + \alpha^2 - 2\alpha \left( \frac{8\pi G_0}{3c_0^2} \varepsilon \right)^{1/2} = \frac{8\pi G_0}{3c_0^2} \left( \varepsilon + \frac{3c_0^2}{8\pi G_0} \alpha^2 - 2\alpha \left( \frac{3c_0^2}{8\pi G_0} \right)^{-1/2} \varepsilon^{1/2} \right) \equiv \frac{8\pi G_0}{3c_0^2} (\varepsilon + \varepsilon_{\alpha e} + \varepsilon_{\alpha m}). \end{aligned} \tag{9}$$

Here, $\varepsilon$ is the composite energy density of matter and radiation. Since the second and the third terms emerge from $\alpha$, we have labeled them accordingly—$\varepsilon_{\alpha e}$ as the $\alpha$-energy density and $\varepsilon_{\alpha m}$ as the $\alpha$-matter energy density. As $\alpha$ turns out to be negative, the contribution of $\alpha$-matter is positive. In terms of the standard definition of the critical density $\varepsilon_{c,0}^S \equiv 3c_0^2 H_0^2 / 8\pi G_0$, Equation (9) becomes

$$H^2 = H_0^2 \left( \Omega_{m,0} a^{-3} f^{-1} + \Omega_{r,0} a^{-4} f^{-2} + \Omega_{\alpha e} + \Omega_{\alpha m,0} a^{-3/2} f^{-1/2} \right). \tag{10}$$

Compared with similar expressions for the $\Lambda$CDM model in a flat Universe

$$H^2 = \frac{8\pi G_0}{3c_0^2} \varepsilon + \frac{\Lambda}{3} = \frac{8\pi G_0}{3c_0^2} \left( \varepsilon + \frac{c_0^2}{8\pi G_0} \Lambda \right) \equiv \frac{8\pi G_0}{3c_0^2} (\varepsilon + \varepsilon_\Lambda) = H_0^2 \left( \Omega_{m,0} a^{-3} + \Omega_{r,0} a^{-4} + \Omega_\Lambda \right), \tag{11}$$

we see that the cosmological constant term $\Lambda/3$ of the $\Lambda$CDM model is replaced by the constant $\alpha^2$ in the CCC model. Additionally, there is an evolutionary term, $2\alpha (8\pi G_0 \varepsilon/3c_0^2)^{1/2}$. This term may be considered a dynamic component of the quintessence (time-varying) dark energy, as dark matter that evolves differently than the baryonic matter or something yet unspecified.

Using Equations (6), Eqauation (9) may be written as

$$H^2 = (H + \alpha)^2 + \alpha^2 - 2\alpha(H + \alpha). \tag{12}$$

This equation can be written directly from Equation (2), but one may have difficulty seeing how it relates to the dark energy of the $\Lambda$CDM model. When divided by $H_0^2$, Equation (12) at the current time becomes

$$1 = \frac{(H_0 + \alpha)^2}{H_0^2} + \frac{\alpha^2}{H_0^2} - \frac{2\alpha(H_0 + \alpha)}{H_0^2} \equiv \Omega_{bm,0} + \Omega_{\alpha e} + \Omega_{\alpha m,0} = \Omega_{bm,0} + \Omega_{\alpha e} + 2\sqrt{\Omega_{\alpha e} \Omega_{bm,0}}. \tag{13}$$

We have labeled the first term as $bm$, as it represents just the baryonic matter in the CCC+TL model [59, 60].

Once we know $\alpha$, we can calculate all the energy density components individually. It is instructive to see how the various energy densities evolve cosmologically with redshift $z$. In CCC+TL (covarying coupling constant + tired light) cosmology [58-60], we define the expanding Universe component of the Hubble constant as $H_X$, which turns out to be 82% of the total Hubble constant by fitting the Pantheon+ data [63, 64]. We must, therefore, use $H_X$ instead of $H_0$ when using equations governing the expanding Universe. It yields $-\alpha/H_X = 0.8$. How the densities evolve with $z$ is shown graphically in Figure 1. In Figure 2, we show the evolution of the ratio of the $\alpha$-matter energy density and the baryonic matter energy density.

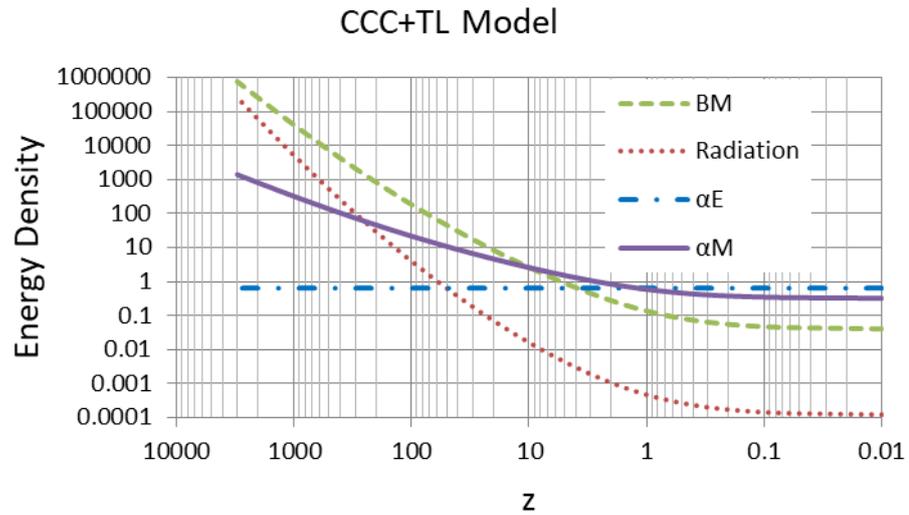

**Figure 1.** Evolution of various energy densities relative to current energy densities in the CCC+TL model plotted against the redshift (BM—baryonic matter, $\alpha E$—$\alpha$ energy, and $\alpha M$—$\alpha$ matter).

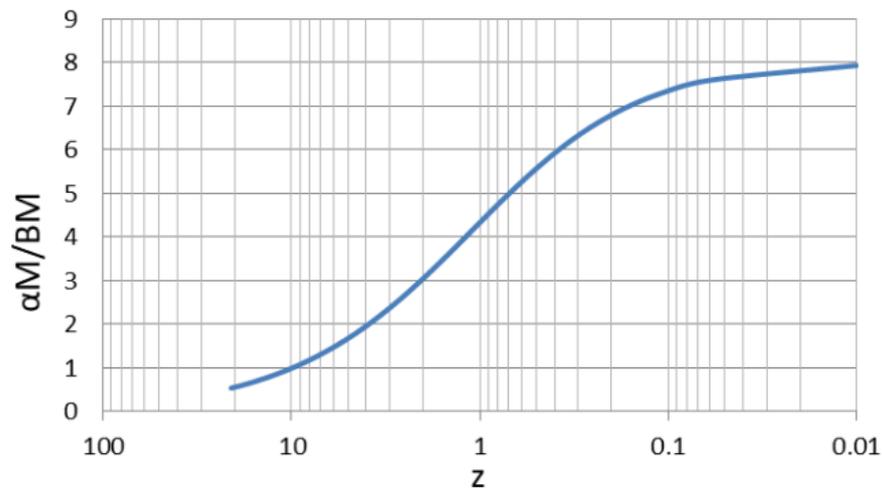

**Figure 2.** Evolution of the ratio of $\alpha$-matter energy density and the baryonic matter energy density shown against the redshift.

Defining a new parameter $X = -\alpha/H_0$, we may write Equation (13) as

$$1 = (1 - X)^2 + X^2 + 2X(1 - X). \tag{14}$$

It shows that $0 \leq X \leq 1$ to satisfy this equation when ensuring that all the energy densities have non-negative values. The first and last terms on the right-hand side of Equations (13) and (14), $\Omega_{bm,0}$ and $\Omega_{\alpha m,0}$ ($= 2\sqrt{\Omega_{\alpha e,0}\Omega_{bm,0}}$), contain baryonic components and thus are directly affected by gravitation. Therefore, X, and consequently $\alpha$, can change locally due to the baryon accretion under gravitational force. It is similar to the energy densities considered constant at *cosmological* scales, such as the critical density, but is drastically different at the *astrophysical* scales of stars and galaxies. In high baryon density regions, only baryons exist, i.e., $X = 0$. At the maximum possible value of $X = 1$, only $\alpha$-energy exists. When $0 < X(r, \theta, \phi) < 1$, all energy densities are present and are responsible for the gravitational pull on any object. Thus, it is the gravitation emerging cumulatively from energy densities $\Omega_{bm,0} + \Omega_{\alpha e} + \Omega_{\alpha m,0}$ ($= 1$) that determines galaxy rotational curves, gravitational lensing due to galaxy clusters, velocity dispersion in galaxy clusters, galaxy formation, etc. However, all these energy densities are relative to the standard critical density ($\varepsilon_{c,0}^S \equiv 3c_0^2 H_0^2/8\pi G_0$) rather than the critical density defined in the CCC model ($\varepsilon_{c,0}^C \equiv 3c_0^2(H_0 + \alpha)^2/8\pi G_0 = (1-X)^2 \varepsilon_{c,0}^S$). Thus, in order to convert relative densities to physical densities in the CCC model, we need to multiply them with $(1-X)^2$. Our focus in this paper is on understanding the galaxy rotation curves in the CCC+TL model.

## 3. Galaxy Rotation Curves

Under spherical galaxy approximation, for simplicity, the deemed observed mass $M_o(R)$, contributed by all the energy densities (as discussed above) at a radius $R$, is related to the Keplerian velocity $V_o(R)$ of a particle by (o-subscript, different from subscript 0, identifies quantities derived from observations)

$$\frac{GM_o(R)}{R} = V_o(R)^2; \quad M_o(R) = 4\pi \int_0^R dr\, \rho(r)\big(1 - X(r)\big)^2 r^2,$$
$$\frac{dM_o(R)}{R^2 dR} = 4\pi \rho(R)\big(1 - X(R)\big)^2. \tag{15}$$

Here, the matter density $\rho(R)$ of the high-density regions of the galaxies is modified by $\big(1 - X(R)\big)^2$ in the low-density regions. When matter density is high (for low values $R < R_t$, where $R_t$ is defined below), $X(R) = 0$. Therefore

$$4\pi \rho_o(R) = \frac{dM_o(R)}{R^2 dR} \qquad R \leq R_t. \tag{16}$$

It can be determined from observed galaxy rotation curves by numerical differentiation. For $R > R_t$ (turn-off radius defined by Daod & Zeki [65]), the density variation is

$$4\pi \rho_o(R) = \frac{dM_o(R)}{R^2 dR} = 4\pi \rho(R_t)(1 - X(R)^2) \qquad R > R_t. \tag{17}$$

Having found $\rho_o(R_t)$ ($\equiv \rho_t$, the turn-off density) using Equation (16), $X(R)$ can now be determined using Equation (17) by keeping $\rho_t$ constant for $R > R_t$. Here, $R_t$ corresponds to the density $\rho_t$ where non-Keplerian rotation dynamics become perceptible. With $X(R)$ known, we use Equation (15) in reverse, using numerical integration while replacing $\big(1 - X(r)\big)^2$ with $\big(1 - X(r)\big)^2 \times \big(1 - X(r)\big)^2$ to find the baryon mass $M_{bX}(R)$ (see Equations (13) and (14) and the corresponding velocity curve $V_{bX}(R)$ and compare $V_{bX}(R)$ with $V_b(R)$, determined by alternative methods represented in the SPARC database [23]).

It should be mentioned that the SPARC data is provided for coarsely spaced radial values with significant uncertainties. This makes it difficult to reliably determine the density using Equation (17) since differentiation is sensitive to the data quality. Fortunately, numerical integration heals this problem substantially.

The question is how do we determine $R_t$? It is not possible to apply the criterion of Daod and Zeki [65] developed for the galaxy NGC3198 to most other galaxies. Additionally, since most SPARC data is at coarsely spaced radial values, it is not possible to select the value that might provide the best fit for a galaxy. A better approach is to use a density-based turn-off parameter $\rho_t$. It provides a more robust galaxy parameter than the turn-off radius and is possibly more relevant in our context. We then find $R_t$ bounds for a galaxy that includes the turn-off density.

Using the density turn-off criterion inspired by Daod and Zeki's [65] radius turn-off criterion, we analyzed the SPARC data for the galaxy NGC3198, as described above, and obtained the curves depicted in Figure 3. Considering that the baryon curve $V_b$ from the SPARC database is an approximation based on multiple observations for the galaxy's bulge, gas, and disk, and our fit $V_{bX}$ is ideally for a spherical galaxy morphology, the two baryon curves are comparable in shape and values.

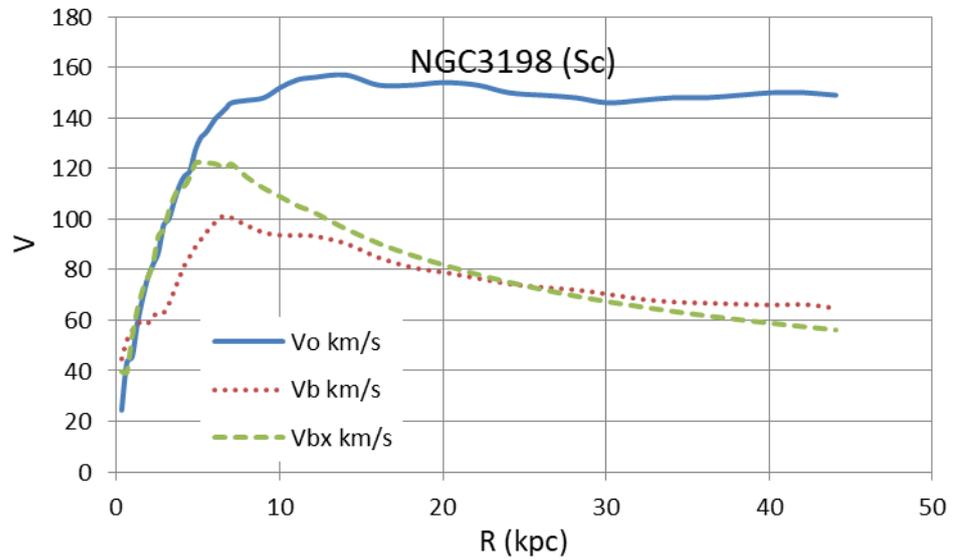

**Figure 3.** Galaxy rotation curves for NGC3198 (Sc) with turn-off density $\rho_t = 2.09 \times 10^{-24}$ g cm$^{-3}$ at the turn-off radius $R_t$ =3.37 kpc. Vo designates the observed rotational velocity curve and Vb the baryon contribution to it, as per the SPARC. VbX label is for the derived baryon matter from the CCC+TL model.

It is natural to ask what makes the density behavior switch from Equation (16) to Equation (17)? From Equations (13) and (14), we may label various density terms as follows: $1 = [(1-X)^2]_{bm} + [X^2]_{ae} + [2X(1-X)]_{am}$. Cosmologically, we have $X = 0.8$ in the CCC+TL model. Matter perturbations grow with time, meaning the first term would become larger with time, i.e., $X$ will become smaller and smaller and approach zero. However, matter accumulation keeps happening under Newtonian gravitation even after $X$ is effectively zero, i.e., when the second and the third terms are effectively zero. This may be considered asymptotic behavior. We may, therefore, define the turn-off density as the density above which the second and third terms are effectively negligible.

The tabulated data in the SPARC database [23] has to be corrected for the mass-to-light ratios for galaxy disks and bulges. We have used nominal values of 0.5 for disks and 0.7 for bulges suggested by McGaugh (email communication; see also [66]) to correct the respective velocities in the SPARC database that uses the mass-to-light ratio of unity for both. Thus, we compute the baryon velocity $V_b$ curves using $V_b = \left(V_{gas}^2 + 0.5V_{disc}^2 + 0.7V_{bul}^2\right)^{1/2}$.

The density profiles $\rho_o$ and $\rho_{bX}$ obtained in the process of computing the $V_{bX}$ curves are plotted in Figure 4. It can be compared with the ideal density profile depicted in Figure 5 for $R \geq R_t$, assuming flat rotation curves for $R > R_t$: $\rho_o = \rho_t(1 - X(R))^2$, $\rho_{bX} = \rho_t(1 - X(R))^4$, $\rho_{\alpha e} = \rho_t(1 - X(R))^2 X(R)^2$, and $\rho_{\alpha m} = 2\rho_t X(R)(1 - X(R))^3$. The profiles of $\rho_o$ and $\rho_{bX}$ in Figure 4 for $R \geq R_t = 3.54$ kpc are similar to the respective profiles of $\rho_0$ and $\rho_{bX}$ in Figure 5, considering that Figure 5 is exhibited with dimensionless axes. Here $X(R) = 1 - R^{-1}$, as derived below for $R \gg R_t$.

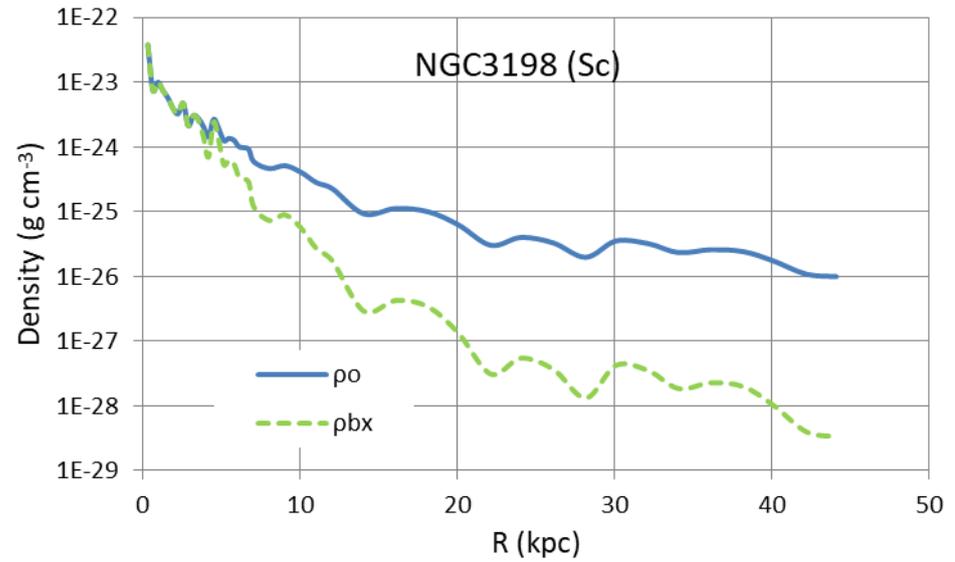

**Figure 4.** Density determined from the galaxy rotation curves for NGC3198 (Sc) with turn-off density $\rho_t = 2.09 \times 10^{-24}$ g cm$^{-3}$ at the turn-off radius $R_t$ =3.37 kpc. Here $\rho_o$ designates the density determined from the observed rotational velocity curve and $\rho_{bX}$ is the baryon density estimated with Equations (15) and (16). The waviness in the curves is an artifact of numerical differentiation (see also Criss and Hofmesiter [67]; they provide density plots for 51 galaxies).

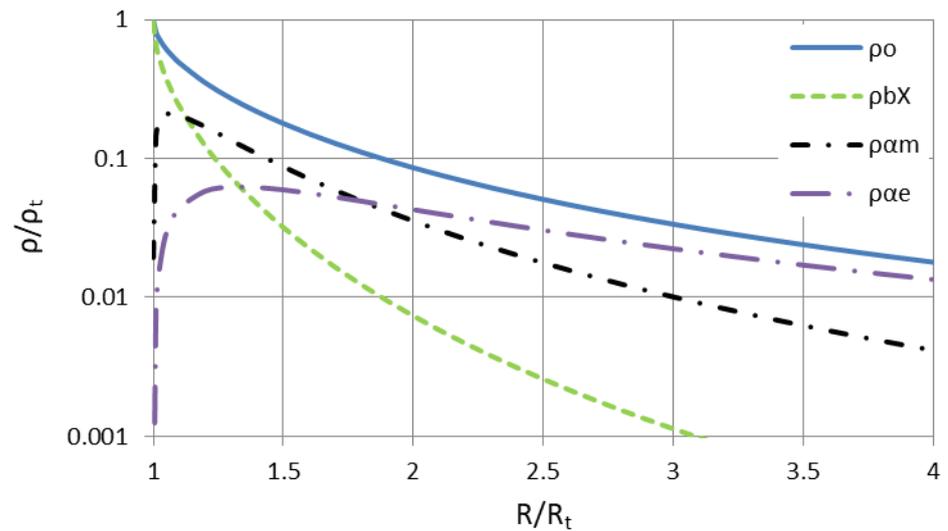

**Figure 5.** Variation in energy densities $\rho_o$, $\rho_{\alpha b}$, $\rho_{\alpha m}$, and $\rho_{\alpha e}$ relative to the turn-off density $\rho_t$ with distance beyond the turn-off distance.

Let us see how the densities $\rho_o$ and $\rho_{bX}$ scale with $R$ when the observed rotation curve $V_0(R)$ is flat, i.e., constant. Then, differentiating $GM_o(R)/R = V_o(R)^2$ of Equation (15), we obtain

$$\frac{GdM_o(R)}{RdR} - \frac{GM_o(R)}{R^2} = 0 \Rightarrow \frac{GdM_o(R)}{R^2 dR} = \frac{GM_o(R)}{R}\left(\frac{1}{R^2}\right) = \frac{V_o(R)^2}{R^2}, \text{ or}$$
$$\frac{dM_o(R)}{R^2 dR} \propto \frac{1}{R^2}. \tag{18}$$

Therefore, from Equation (17)

$$4\pi\rho_o(R) = 4\pi\rho(R_t)(1-X(R))^2 = \frac{V_o(R)^2}{GR^2} \Rightarrow (1-X(R))^2 = \frac{1}{R^2} \Rightarrow X(R) = 1 - R^{-1}, \tag{19}$$

and we obtain $\rho_o(R) \sim 1/R^2$ beyond the turn-off radius, when $\rho$ is constant $\rho_t$. Since $\rho_{bX} \sim (1-X(R))^4$, we get $\rho_{bX} \sim 1/R^4$ for $R \gg R_t$, similar to Hernquist's model [68] in which density falls as $R^{-4}$ beyond a characteristic radius [69]. Recall that we have used a spherical approximation for galaxy morphology. The density of galaxies, scaling as $R^{-2}$, was obtained for 51 galaxies without dark matter by Criss and Hofmeister[67], considering their oblate geometry (see also Hofmeister et al. [28]).

We have randomly selected six more galaxies from the SPARC database to see how well our approach determines the baryon velocity profile from the observed velocity profile and compare the former with the approximate baryon velocity profile given in SPARC. The first one we analyzed is D631-7 (Im), which has a galaxy classification designated by Im. Its velocity curves are shown in Figure 6. We used the turn-off density given by $\rho_t = 2.79 \times 10^{-24}$ g cm$^{-3}$, giving the turn-off radius $R_t = 0.35$ kpc. It should be mentioned that $V_{bX}$ determination is sensitive to the $\rho_t$ value, and hence the $R_t$ value for a galaxy.

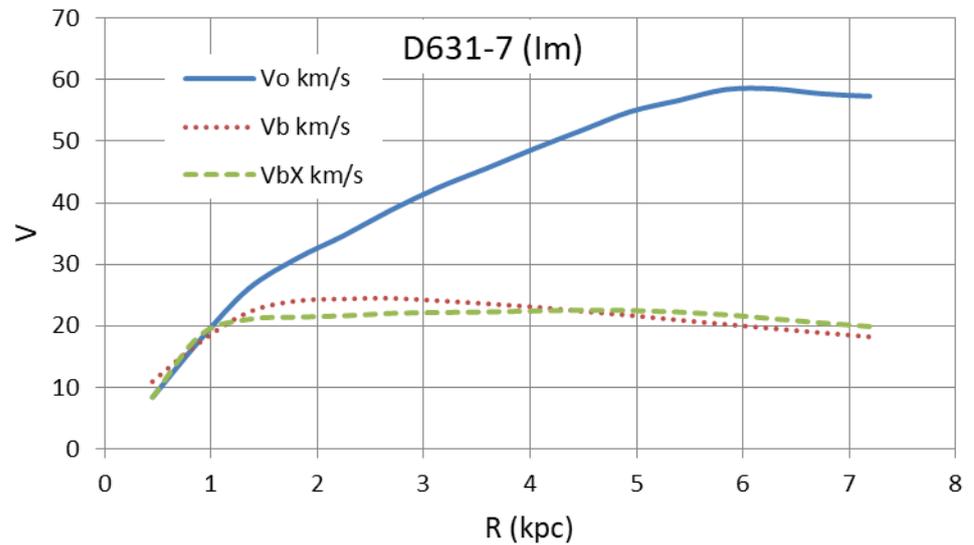

**Figure 6.** Galaxy rotation curves for D631-7 (Im) with turn-off density given by $\rho_t = 2.79 \times 10^{-24}$ g cm$^{-3}$, giving the turn-off radius $R_t = 0.35$ kpc. The label description is the same as in Figure 3.

Figures 6–11 depict the velocity curves for six other galaxies. As the data in the SPARC database [23] is given at sparsely populated $R$ values, the turn-off radius $R_t$ for a galaxy, which gives the best visual fit of $V_{bX}$ with $V_b$ for the galaxy, does not correspond to a tabulated $R$ value. Instead, we find the turn-off density $\rho_t$ and bound for the $R_t$ from the tabulated $R$ values.

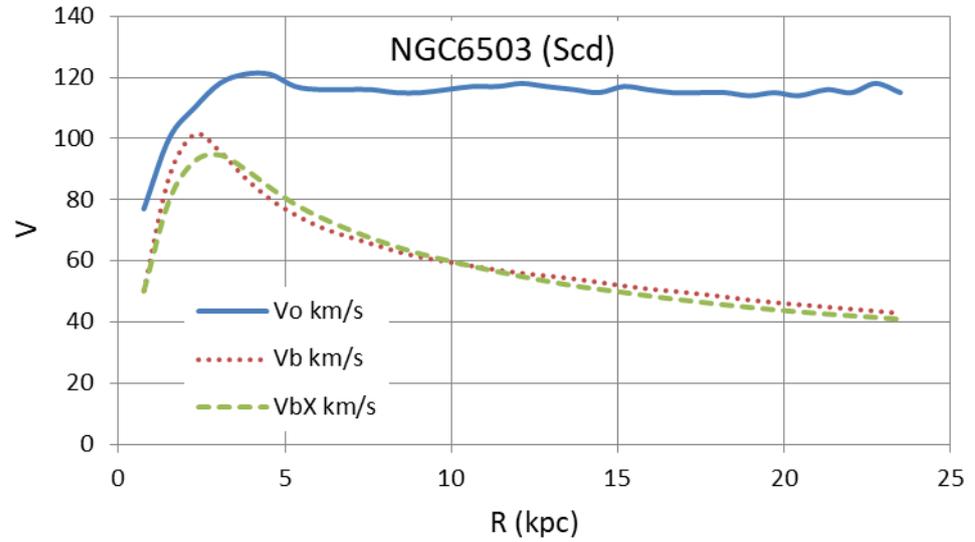

**Figure 7.** Galaxy rotation curves for NGC6503 (Scd), with turn-off density given by $\rho_t = 5.17 \times 10^{-24}$ g cm$^{-3}$, giving the turn-off radius $R_t = 1.9$ kpc. The label description is the same as in Figure 3.

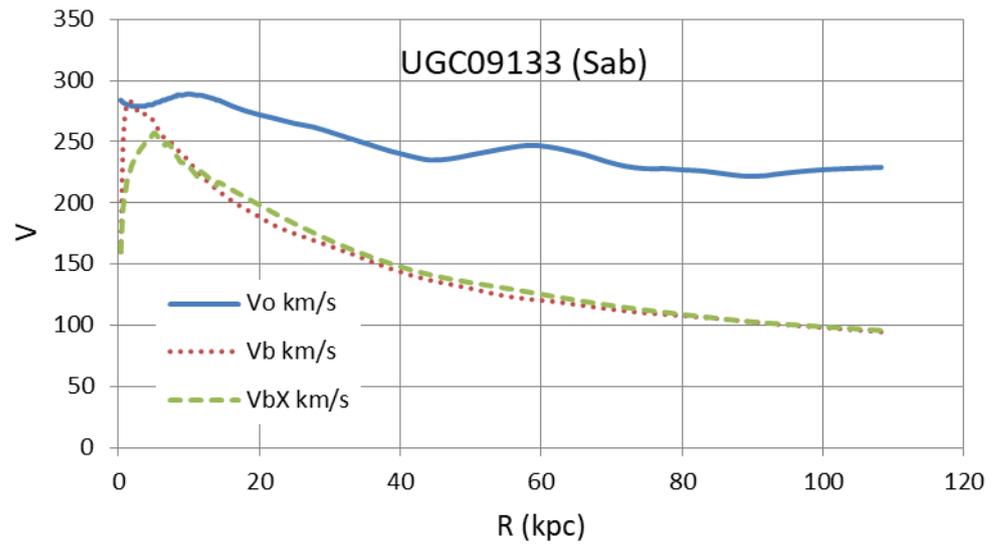

**Figure 8.** Galaxy rotation curves for UGC09133 (Sab), with turn-off density given by $\rho_t = 3.98 \times 10^{-24}$ g cm$^{-3}$, giving the turn-off radius $R_t = 4.9$ kpc. The label description is the same as in Figure 3.

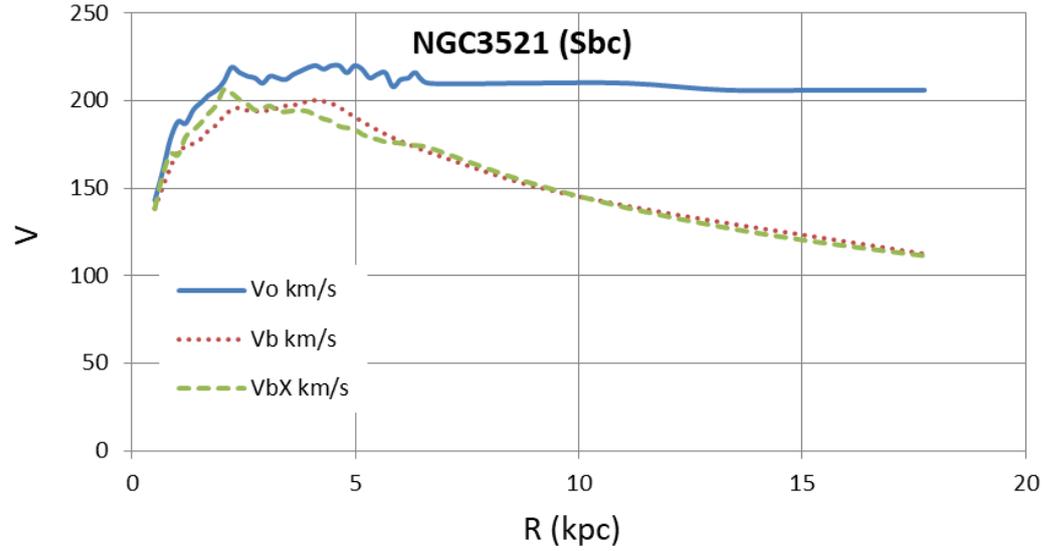

**Figure 9.** Galaxy rotation curves for NGC3521 (Sbc), with turn-off density given by $\rho_t = 7.96 \times 10^{-24}$ g cm$^{-3}$, giving the turn-off radius $R_t = 2.15$ kpc. The label description is the same as in Figure 3.

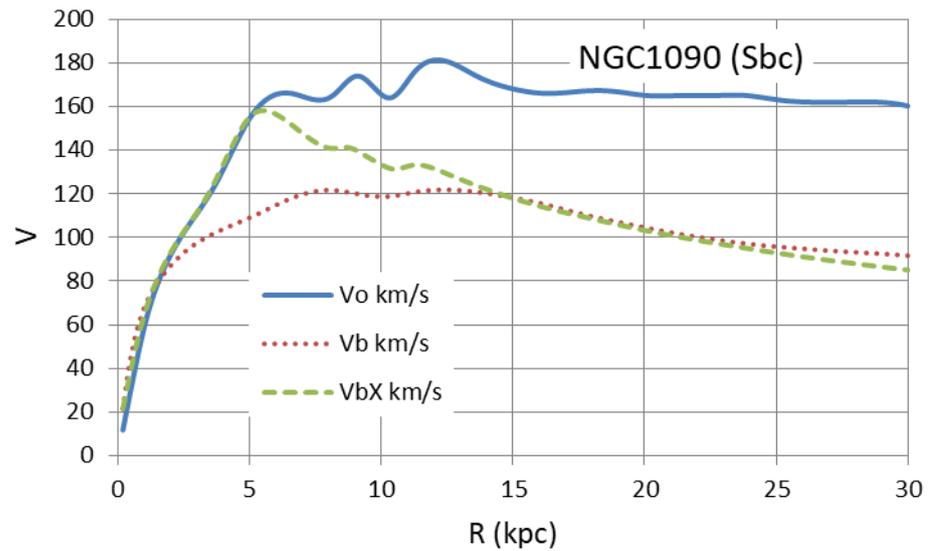

**Figure 10.** Galaxy rotation curves for NGC1090 (Sbc), with turn-off density given by $\rho_t = 2.23 \times 10^{-24}$ g cm$^{-3}$, giving the turn-off radius $R_t = 4.7$ kpc. The label description is the same as in Figure 3. We used a spline interpolated grid from irregularly spaced data for this galaxy for better numerical differentiation.

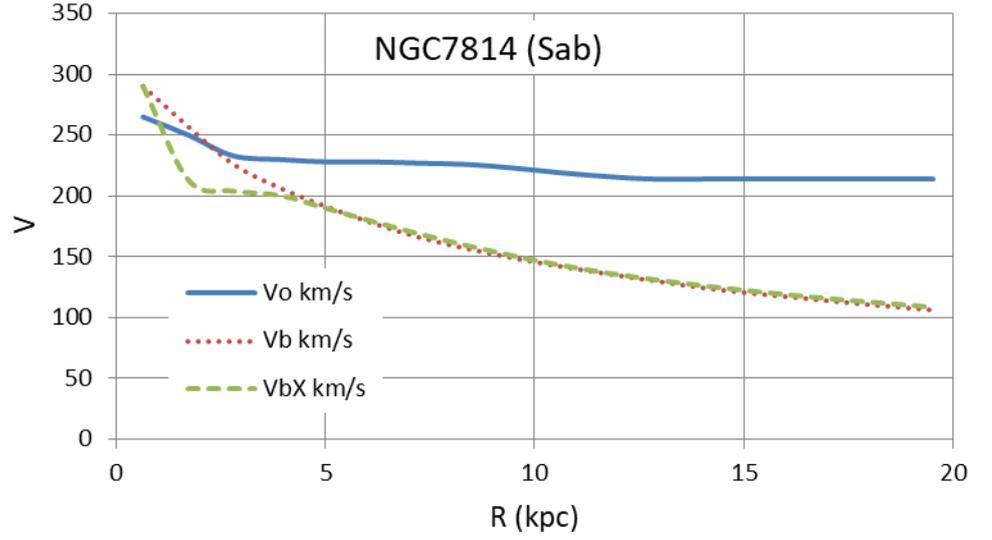

**Figure 11.** Galaxy rotation curves for NGC7814 (Sab), with turn-off density given by $\rho_t = 5.01 \times 10^{-24}$ g cm$^{-3}$, giving the turn-off radius $R_t = 2.8$ kpc. The label description is the same as in Figure 3.

We wish to emphasize that our method for fitting the SPARC data is different than that which is used normally. The normal method is to estimate the baryonic matter of a galaxy from different observations and fit the observed rotation curves by adding dark matter determined from some model to the baryonic matter. Our method uses the observed rotation curves to determine the parameter $X(R)$ and then use this $X(R)$ to fit the baryonic rotation curve with one free parameter and compare it with a composite baryonic matter curve using the SPARC database. We show that the predicted baryonic rotation curves are close to those from the database. This is a proof-of-concept paper with spherical approximation for galaxies. Most studies have used a more realist approximation for galaxy shapes.

The fit data for the galaxies is presented in Table 1, including the remarks about their visual quality of fits. We notice that while the turn-off radius $R_t$ spread is over a factor of about eight, the turn-off density $\rho_t$ spread is only over a factor of about four. Thus, we consider the turn-off density to be a more robust fit parameter. The fact that these parameters change from galaxy to galaxy is not surprising considering our very simple spherical model; galaxies are far from being spherical and have different morphologies. It is worth mentioning that the turn-off acceleration $a_t \equiv V_{flat}^2/R_t$, where $R_t$ is derived from the turn-off density $\rho_t$ for the CCC models, turns out to be about the same order as the Milgromian (MOND) acceleration $a_0 = 1.2 \times 10^{-8}$ cm s$^{-2}$. This is just for comparison purposes, since CCC models have no dependence on MOND; $a_t$ is not used in our work in this paper, unlike the works focusing on MOND [e.g., 23, 56, 69, 70]). It is worth emphasizing that we have used only one free parameter $\rho_t$ to fit the galaxy rotation curves. The *cosmological parameters* $H_0$ and $\alpha$, derived in earlier papers [e.g., 58] by fitting the Pantheon+ data, are not relevant for fitting the astrophysical SPARC data.

**Table 1.** Turn-off radius $R_t$ and turn-off densities $\rho_t$ for the galaxies in this work. The visual quality of fit is stated under the remarks column. Other parameters shown do not have direct relevance to our findings but are interesting for comparison purposes. Size here is the effective radius $R_{eff}$ at 3.6 μm, as defined in the SPARC database.

| Galaxy | Type | Distance | Size | $R_t$ | $\rho_t$ | $a_t$ | Remarks |
|---|---|---|---|---|---|---|---|
| | | Mpc | kpc | kpc | $10^{-24}$ g cm$^{-3}$ | $10^{-8}$ cm s$^{-2}$ | |
| D631-7 A | Im | 7.72 | 1.22 | 0.35 | 2.79 | 3.10 | Good overall fit |
| NGC1090 | Sbc | 37.00 | 6.36 | 4.7 | 2.23 | 1.76 | Not a good fit for $R \leq 15$ kpc |
| NGC3198 | Sc | 13.80 | 5.84 | 3.37 | 2.09 | 2.15 | OK overall fit |
| NGC3521 | Sbc | 7.70 | 2.45 | 2.15 | 7.96 | 6.43 | Good fit |
| NGC6503 | Scd | 6.26 | 1.62 | 1.9 | 5.17 | 2.25 | Good fit |
| NGC7814 | Sab | 14.40 | 2.08 | 2.8 | 5.01 | 5.33 | Good fit |
| UGC09133 | Sab | 57.10 | 5.92 | 4.9 | 3.98 | 3.33 | Good overall fit |

## 4. Discussion

The primary purpose of this paper is to explore whether the covarying coupling constants in cosmology are *astrophysically* consistent. One of the most essential consistency tests is the gravitational rotation curves, since CCC cosmology has no dark matter per se. However, Friedmann equations may be written in a way that the covarying coupling constant parameter $\alpha$ results in terms that appear to behave like dark matter and dark energy (Equation (9)). Since these terms emerge from $\alpha$, we have labeled them as $\alpha$-matter and $\alpha$-energy, the former being twice the square root of the product of the standard matter term and the $\alpha$-energy term (Equation (13)).

While cosmologically $\alpha$ is a constant determined for the homogenous and *isotropic* Universe by fitting Pantheon+ data, it can vary locally within its limit of 0 to $-H_X$ due to extreme *anisotropy* of the matter distribution. By defining $X = -\alpha/H_X$, we have $0 \leq X \leq 1$ (Equation (14)) for all energy densities to be positive. Recall that $H_0$ is replaced by $H_X$ in the hybrid CCC+TL model that includes tired light. By allowing the variation of $X$ such that only baryonic matter exists in a galaxy at small radii as observed (highly luminous matter density region) below a turn-off baryon density $\rho_t$ (corresponding turn-off radius being $R_t$) and no baryonic matter at $R \gg R_t$, and by fitting the observed galaxy rotation velocity curve using the baryonic matter plus the $\alpha$-matter and $\alpha$-energy, we can determine $X(R)$ as described in the previous section (Equations (15)–(17)). Once $X(R)$ is known, we can determine, individually, the baryonic matter term, the $\alpha$-matter term, and the $\alpha$-energy term. It is worth reiterating that since the $\alpha$-matter term involves $\alpha$-energy, it may be equated to the dynamic component of dark energy, i.e., to the quintessence (time-varying dark energy).

We also notice that when we constrain the observed rotation curves to be flat at a large $R$, the total density scales a $R^{-2}$, as expected. However, the baryonic density scales are $R^{-4}$ for $R \gg R_t$ (see Equations (18) and (19)), the same as in the Hernquist model (Hernquist 1990)[68]. These findings are based on a simplified spherical approximation for galaxy morphology and need correcting. Additionally, our computations involve numerical differentiations and integrations. Since the data points are sparse and often unevenly spaced, the computations are prone to significant numerical errors.

In the CCC models, galaxy dynamics depend not only on the matter density, but also on the $\alpha$-energy density and the $\alpha$-matter density. As shown in Figure 1 and Equation (10), they scale differently with the scale factor. Thus, the greatly reduced contribution of $\alpha$-energy density and the $\alpha$-matter density at a higher redshift to even a virialized system cannot be ignored. For example, the ratio of $\alpha$-matter density relative to the baryonic matter density decreases with the increasing redshift (e.g., from eight at $z = 0$ to three at $z = 2$; Figure 2), while the $\alpha$-energy remains constant, one should expect the rotation curves would approach baryonic, i.e., Keplerian, for high redshift galaxies, as observed [e.g., 2, 5]. It will be interesting to analyze the rotation velocity curve of the REBELS-25 galaxy ($z = 7.31$, where the ratio decreases to 1.2), the earliest disk galaxy observed [71]. However, currently, there are very limited data points to make a reliable analysis. It should be noted that the $R$ values in the paper are determined using the standard ΛCDM model from the angular size measurements; they will increase by a factor of ≈7 in the CCC+TL model [58] at the REBELS-25's redshift of $z$ = 7.31. No such correction is needed for the velocities as they are measured using the Doppler effect.

How would the galaxy formation be affected in the CCC approach? As the redshift increases, the density ratio of $\alpha$-matter to baryonic matter decreases for a given $\alpha$. However, as discussed above, $\alpha$ can vary locally at any redshift. The ratio is then $2X/(1 - X)$, moderated by the ratio in Figure 2. As $X \to 1$ at $R \gg R_t$, the $\alpha$-matter dominates over baryonic matter, and cumulatively with $\alpha$-energy provides the potential well for baryonic matter to fall in for the galaxy formation. The $\alpha$-matter and $\alpha$-energy thus behave like dark matter in assisting the galaxy formation.

Several factors can determine how X varies in a galaxy, such as its morphology, formation and evolution history, tidal effects, mergers, etc. The shape of its observed rotation curve could be quite varied—rising, flat, declining, etc. Since there is a great deal of uncertainty in a galaxy's baryon mass distribution, it appears appropriate to start with the observed rotation curves and derive the baryon mass distribution as we have achieved above. One does not observe the rotation curves corresponding to a galaxy's bulge, disk, gas, etc. One estimates the mass distribution for each component based on specific models and then converts them into rotation curves for each if the same were observable independently.

Superficially, one could see that the Tully–Fisher relation is easily satisfied in the CCC approach. For a given geometry, mass is proportional to density. Thus, from Equation (19), for observed rotation curves, $(1 - X(R))^2 \propto V(R)^2$ for $R > R_t$. Since baryonic mass is $M_B \propto (1 - X)^4$, we obtain $M_B \propto V^4$, which is compliant with BTFR (Baryonic Tully-Fisher Relation) [70].

Next, we should consider the role of $\alpha$-matter and $\alpha$-energy in the galaxy clusters. To realistically estimate it, we should determine the mass density distribution around each galaxy at its location in the cluster and superimpose them for all the galaxies in a cluster. Nevertheless, we decided to treat a galaxy cluster similarly to a galaxy to determine its density profiles and calculate the baryonic density curve to compare it with that in the original paper (Mandelbaum et al. [72], Figure 1). It is depicted in Figure 12 and shows a good visual fit when the turn-off density is set $\rho_t = 2.0 \times 10^{-24}$ g cm$^{-3}$. It turns out to be in the same range as the turn-off density for galaxies in Table 1. It is interesting to compare Figures 4, 5, and 12 while keeping in mind that the axes are ratios in Figure 5 and the $y$-axis unit in Figure 12 is different from the $y$-axis unit in Figure 4. Additionally, one should expect the effect of $\alpha$-matter to decline with the increasing redshift for a galaxy cluster, as in Figure 2.

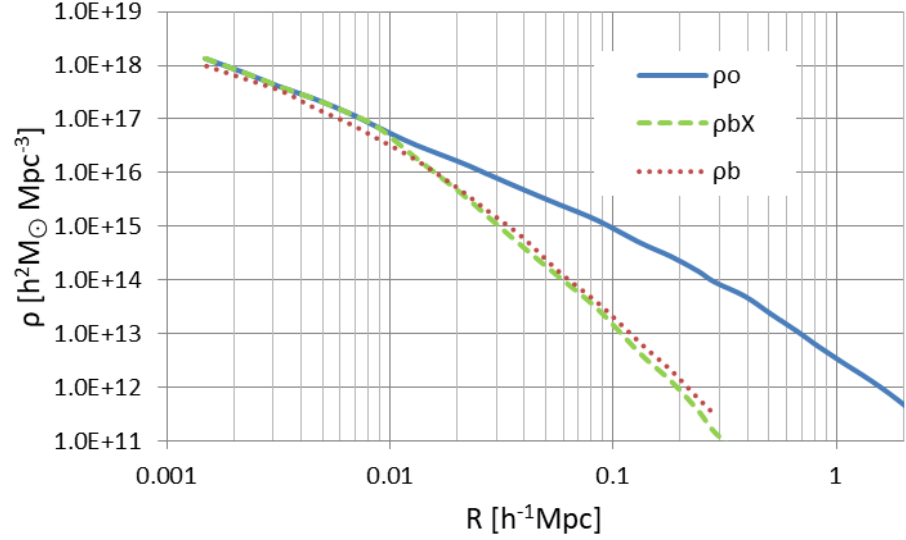

**Figure 12.** Density profile $\varrho(R)$ for a typical galaxy cluster. The total density profile $\rho_o$ (blue line) and baryonic density profile $\rho_b$ (red dotted line) are from Mandelbaum et al. (2006). The reduced Hubble constant is $h$. The density $\rho_{bX}$ is calculated using $\rho_o$, similarly to our calculations for galaxy rotation curves with $\rho_t = 5.74 \times 10^{16}\, h^2 M_\odot\, \text{Mpc}^{-3} = 2.0 \times 10^{-24}\, \text{g cm}^{-3}$ (assuming $h = 0.72$ for the standard cosmology).

It is worth reiterating that $\alpha$-matter is not dark matter; it is the effect of covarying coupling constants and nothing physical, contrary to the physical dark matter of the $\Lambda$CDM cosmology. Along with baryonic matter, the $R$-dependence of $\alpha$-matter and $\alpha$-energy affects gravitational lensing, galaxy velocity dispersion, etc., in the same way as dark matter in the $\Lambda$CDM model. Thus, it is the cumulative gravitational potential well of baryons, $\alpha$-matter, and $\alpha$-energy that determines the galaxy rotation curves and other gravitational dynamics in galaxies and galaxy clusters, galaxy formations and evolution, etc.

The morphologies of galaxies are highly varied. Treating them as spherical is an oversimplification. One must allow for such variation to better fit the observed rotation curves.

We should now address concerns about the general covariance of the CCC approach. The CCC concept was inspired in the framework presented by Costa et al. [73]. Therein, the covarying physical couplings were derived from an action integral (i.e., a Lagrangian density), Einstein-like field equations were built deductively, and general constraints relating the simultaneously varying coupling were determined (see also [74]). Modified Friedmann equations stem from the extended Einstein field equations when they are specified for the homogeneous and isotropic Universe. The concern about picking a special time coordinate, thus potentially breaking the general covariance, is addressed in our papers [62, 75]. We show in these papers how to ensure a general covariance even when a varying speed of light (VSL) participates in the time sector of the line element. We also discuss energy–momentum conservation thoroughly. Therefore, the standard notion of energy–momentum conservation has to be extended to encompass the varying couplings. This feature is present in other VSL proposals, such as by Albrecht and Magueijo[76]. It is also present in Brans–Dicke theory and other scalar–tensor theories of gravity, as discussed in Chapter 2 of the book by Faraoni[77]. Equation (4) here is a manifestation of this extended conservation law of the energy–momentum tensor for a perfect fluid. In standard cosmology, $\alpha$ is identically zero, and we recover the standard result.

Another general concern relates to the tired light included in the CCC+TL model. We believe that tired light effect exists in parallel to the expansion of the Universe. We have dealt with this in earlier cited papers in detail [58-60]. Since the distance traveled by the photon is the same, this fact is used to correlate the two causes of the observed redshift without requiring an additional parameter. The time dilation concern about the tired light was discussed in Gupta [60]. The fact that the CCC+TL model fits the Pantheon+ data as well as the ΛCDM model shows that Tolman brightness concern is accommodated in the CCC+TL model. We do not suggest scattering to be the cause of tired light. It is currently unknown and the subject of ongoing research.

With the turn-off acceleration $a_t$ being so close to the MOND non-Newtonian acceleration $a_0$, it is natural to ask if the CCC model is related to MOND. While the cause of MOND $a_0$ is unknown and unrelated to cosmology, $\alpha$-matter and $\alpha$-energy are clearly related to the variation in coupling constants. Since $a_t$ values from the CCC model, derived from $\rho_t$ as discussed above, are close to MOND's $a_0$, one may conclude that the MOND transition of acceleration from Newtonian to non-Newtonian may be caused by varying coupling constants, i.e., MOND may be a manifestation of the CCC effect.

## 5. Conclusions

We conclude the following:

1. Dark matter and dark energy of the ΛCDM model may be considered emerging from the weakening of the forces of nature in an expanding Universe determined by a dimensionless covarying coupling constant function. For simplicity, we have defined the function as $f(t) = \exp(\alpha(t - t_0)]$, with $\alpha$ as a constant.
2. The observed galaxy rotation curves can be used to derive the baryonic matter distribution in a galaxy. One may not need to bring in new physics, such as modified gravity, Modified Newtonian Dynamics, etc., to explain the observed flat rotation curves of galaxies. The CCC approach is thus *cosmologically* and *astrophysically* consistent.
3. The emergent $\alpha$-matter and $\alpha$-energy can, in principle, replace dark matter in galaxy clusters and assist in galaxy formation. However, unlike in the ΛCDM cosmology, $\alpha$-matter is not something physical.
4. The simplified spherical galaxy approach we have used in this paper needs correction to account for the observed galaxy morphologies to obtain a better data fit.

**Funding:** This research received no external funding.

**Data Availability Statement:** References have been provided for the data used in this work.

**Acknowledgments:** The author is thankful to Professor Stacy McGaugh and Professor Rodrigo Cuzinatto for the communications that helped shape this work. He is grateful to the three reviewers of the paper for their very constructive critical comments that resulted in greatly improving its contents, quality, and clarity.

**Conflicts of Interest:** The authors declare no conflict of interest.